\documentclass[letter,twocolumn,amsmath,amssymb,superscriptaddress]{jpsj3}
\usepackage{txfonts,amsmath,amssymb}
\usepackage{bm}
\usepackage{braket}

\def\runauthor{A. Ozawa, K. Nomura}

\title{Two-orbital effective model for magnetic Weyl semimetal in Kagome-lattice shandite}

\author{Akihiro Ozawa$^1$, 
Kentaro Nomura$^{1,2}$\thanks{nomura@imr.tohoku.ac.jp}
}
\inst{$^1$Institute for Materials Research, Tohoku University, Katahira, Aoba-ku, Sendai 980-8577\par
$^2$Center for Spintronics Research Network, Tohoku University, Katahira, Aoba-ku, Sendai 980-8577
}

\kword{%
Weyl semimetal, anomalous Hall effect, Kagome lattice, shandite, ferromagnetism
}

\abst{
We construct a two-orbital effective model for a ferromagnetic Kagome-lattice shandite, $\rm{{Co}_3{Sn}_2{S}_2}$, a candidate material of magnetic Weyl semimetals, by considering one $d$ orbital from Co, and one $p$ orbital from interlayer Sn.
The energy spectrum near the Fermi level, and the configurations of the Weyl points, computed by using our model, are similar to those obtained by first principle calculations.
We show also that nodal rings appear even with spin-orbit coupling when the magnetization points in-plane direction.
Additionally, magnetic properties of $\rm{{Co}_3{Sn}_2{S}_2}$ and other shandite materials are discussed. 

}

\begin{document}
\maketitle

%\section{Introduction}

{\it Introduction}---
Weyl semimetals are gapless semiconductors with non-degenerate point-nodes called Weyl points\cite{Wan2011,Burkov2011,Armitage2018}. 
These nodes generate a fictitious magnetic field, the Berry curvature\cite{Xiao2010}, in momentum space. 
Weyl semimetals with magnetic order, magnetic Weyl semimetals, attract attention because of the nontrivial charge-spin coupling, represented by the large anomalous Hall effect \cite{Nagaosa2010}.
Additionally, some electromagnetic responses such as the charge induced spin torque\cite{Nomura2015,Kurebayashi2016} and electric-field-driven domain wall motion\cite{Araki2016,Kurebayashi2019} are theoretically predicted.
These phenomena suggest a potential to implement highly efficient magnetic devices by using the topological characters of electronic states.
As candidates, some materials such as $\rm{{Mn}_3{Sn}}$\cite{Chen2014,Kubler2014,Nakatsuji2015,Kuroda2017,Suzuki2017,Ito2017,Jianpeng2017,Yang2017} and Heusler alloys \cite{Zhijun2016,Guoqing2016} have been studied. 
However, these materials have large Fermi surfaces in addition to the Weyl points.
This metallicity may suppress functionalities of Weyl semimetals mentioned above by the screening effect.
In order to realize these functionalities, it is important to find magnetic Weyl semimetals with small Fermi surfaces.\par
Very recently, it was suggested that a ferromagnetic Kagome-lattice shandite $\rm{{Co}_3{Sn}_2{S}_2}$ is a strong candidate of the magnetic Weyl semimetal from first principle calculations and experiments\cite{Enke2018,Qiunan2018,Qi2018}. 
This material possesses a relatively large anomalous Hall angle $\theta_{\rm{AHE}}={\it{\sigma}}_{\rm{AHE}}/{\it{\sigma}}_{xx} \approx \rm{20} \%$ which is much larger than those of other candidates such as $\rm{Mn}_3{Sn}~(\theta_{AHA} \approx 3 \%)$\cite{Nakatsuji2015,Enke2018} . 
This indicates the semimetallic character with a small longitudinal conductivity $\sigma_{xx}$ and small Fermi surfaces, suggesting an ideal magnetic Weyl semimetal.
According to first principle calculations\cite{Enke2018,Qiunan2018}, in the absence of spin-orbit coupling, nodal rings appear near the Fermi level, while spin-orbit coupling opens energy gaps on the nodal rings except at some points, the Weyl points.\par  
Utilizing $\rm{{Co}_3{Sn}_2{S}_2}$, we expect electromagnetic functionalities of Weyl semimetals.
However, it is difficult to study the magnetic response using first principle calculations, because we have to introduce  magnetic field via the Peierls phase of electrons that causes a huge matrix of the Hamiltonian with many orbitals. 
Therefore, it is desirable to construct a minimal model describing the low energy excitations with only a few orbitals.\par
In this work, we construct an effective two-orbital tight-binding model by using one of the $d$ orbitals from Co and one of the $p$ orbitals from interlayer Sn. 
We show that configurations of nodal rings and the Weyl points in the Brillouin zone in our model are similar to those in first principle calculations\cite{Enke2018,Qiunan2018}. 
Additionally, we show that nodal rings appear even with spin-orbit coupling when the magnetization points in-plane direction.
We discuss the origin of magnetism in shandite materials and magnetic anisotropy.

%\section{Model Hamiltonian}

\begin{figure}[t]
\includegraphics[width=1\hsize]{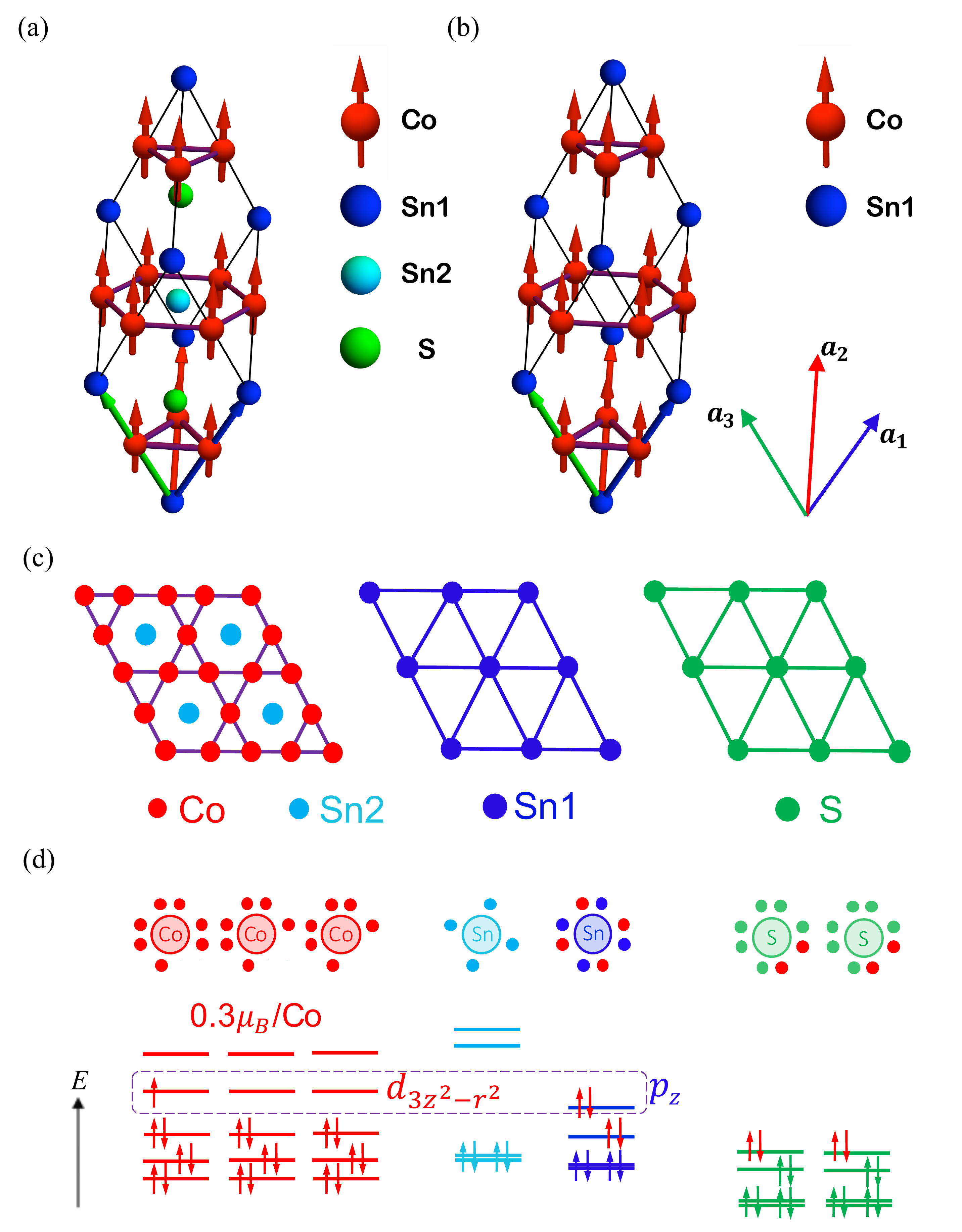}
\vspace{-7mm}
\caption{
(a)~Original unit cell of Co-shandite and 
(b)~unit cell of our model. Co is responsible for ferromagnetic order. 
(c)~Each layer of Co-shandite. Kagome layer contains Co atoms and Sn2 atoms. 
Those Kagome layers sandwich two types of triangle layers of Sn1 atoms and S atoms.
(d)~The energy relation and occupied electrons assumed in our model.

}	
\label{fig:electrons}
\end{figure}

{\it Model Hamiltonian}---
In the following, we construct an effective tight-binding model Hamiltonian by considering the crystal field splitting and relevant orbitals close to the Fermi level of the system.
Figures~\ref{fig:electrons}(a) and \ref{fig:electrons}(b) compare the original unit cell of Co$_3$Sn$_2$S$_2$ and the unit cell of our model.
Co$_3$Sn$_2$S$_2$ consists of primitive rhombohedral unit cells including three Co atoms, two Sn atoms, and two S atoms as shown in Fig.~\ref{fig:electrons}(a).
One layer has the Kagome lattice structure of Co atoms with Sn at the center of hexagons as shown in Fig.~\ref{fig:electrons}(c).
We refer to it as Sn2 to distinguish another Sn atom, Sn1, which form a triangular lattice. 
There are also two layers of triangular lattices of S atoms.
We assume that the crystal field splitting in shandite is as shown in Fig.~\ref{fig:electrons}(d).
Splitting energies in the Kagome layer are larger than those in the triangular lattices of Sn and S atoms because the interatomic distance in the Kagome lattice is shorter than those in the triangular lattices.
In this crystal structure, the five-fold degeneracy of Co's $d$ orbitals is lifted, although the energy-level relationship is not clear.
A neutral Co atom, Sn atom, and S atom have 9, 4, and 6 valence electrons, respectively.
When the crystal field splitting energies are large enough compared to the Hund coupling energies, low spin states are favored; low energy orbitals are occupied by the valence electrons.
The electron's configuration is assumed as shown in Fig.~\ref{fig:electrons}(d), where the fourth of five $d$ orbitals of Co atoms are partially occupied;
there is one electron in $3\times2=6$ states on three Co atoms per unit cell.
 When spins are fully polarized this configuration is consistent with the magnetic moment $M_{\rm{Co}} \approx 0.3\mu_B /\rm{Co}$ suggested by first principle calculations\cite{Enke2018} and experiments\cite{Kubodera2006,Paz2009,Qi2018}.
We assume that $d_{3z^2-r^2}$ is the partially occupied orbital and that the occupied $p_z$ orbital of Sn1 atom is close
to the Fermi level. On the other hand, all other orbitals are far from the Fermi level and thus neglected in the following.
In this model, there are three electrons in eight bands, 6 bands from three Co atoms and 2 bands from one Sn1 atom.
The Fermi level is determined by this 3/8 filling condition.
The unit cell of our model has a rhombohedral lattice structure as the original one.
The primitive translation vectors are $\bm{a}_1=(\frac{a} {2},0,c)$, $\bm{a}_2=(-\frac{a}{4},\frac{ \sqrt{3}a}{4},c)$, $\bm{a}_3=(-\frac{a}{4},-\frac{\sqrt{3}a}{4},c)$ as shown in Fig.~\ref{fig:electrons}(b). 
In the following, we set $c=\frac{\sqrt{3}a}{2}$ for simplicity.
In our model, the unit cell includes three Co atoms on the Kagome lattice and one Sn1 atom on the triangular lattice.

Our effective model Hamiltonian consists of three terms, 
\begin{align}
H&=H_{\text{d-p}}+H_{\rm{exc}}+H_{\rm{so}},
\label{eq:hamiltonian}
\end{align}
where  $H_{\text{d-p}}$ is the hopping term, $H_{\rm{exc}}$ is the exchange coupling term, and $H_{\rm{so}}$ is the spin-orbit coupling term. 
Detailed explanations of each term are given in the following.\par
We start with the hopping term $H_{\text{d-p}}$.
We consider the first and second-nearest-neighbor hopping, $t_1$ and $t_2$, in the Kagome layer, inter-Kagome-layer hopping, $t_{\rm{z}}$, and $dp$ hybridization $t_{\rm{dp}}$. 
We neglect hopping between Sn atoms because the interatomic distance is longer than others. 
 $H_{\text{d-p}}$ is written as follows,
 \begin{align}
 H_{\text{d-p}} &= -\sum_{ij\sigma}t_{ij} d^{\dagger}_{i\sigma} d_{j\sigma} ^{}
 -\sum_{ij\sigma}(t^{\rm{dp}}_{ij} d^{\dagger}_{i\sigma} p_{j\sigma}^{}+t^{\rm{dp}}_{ij} p^{\dagger}_{i\sigma} d_{j\sigma}^{})
 \nonumber \\
 &\qquad\qquad + \epsilon_{\rm{p}} \sum_{i\sigma}  p^{\dagger}_{i\sigma} p_{i\sigma}.
 \end{align}
 Here $d_{i\sigma}$ and $p_{j\sigma}$ are the annihilation operators of $d$ electrons on the Kagome lattice and $p$ electrons on the triangular lattice.
 $t_{ij}$ describes the hopping between Co sites and is either of $t_1$, $t_2$, $t_{\rm{z}}$ or zero, depending on the relative position.
 $t^{\rm{dp}}_{ij}=t_{\rm{dp}}$ between nearest Co and Sn1 sites, otherwise $t^{\rm{dp}}_{ij}=0$. $\epsilon_{\rm{p}}$ is the energy difference between $p$ orbital and $d$ orbital.\par
 
  $H_{\rm{exc}}$ describes the ferromagnetic ordering derived from the onsite Hubbard coupling within the mean field approximation\cite{Jurgen2017}. 
The Hamiltonian is given in the following form,
\begin{align}
 H_{\rm{exc}}=-J \sum_{i\sigma\sigma'} \bm{m} \cdot d^{\dagger}_{i,\sigma} \bm{\sigma}_{\sigma\sigma'} d_{i,\sigma'}.
\end{align}
Here,  we neglect the onsite Coulomb energy of $p$ orbital of Sn1 because that of $p$ electrons is smaller than that of $d$ electrons.
In the mean field theory $\bm{m}$ is determined self-consistently. 
On the other hand, in the following, we set  the strength of $J|\bm{m}|$ by comparing to the first principle calculations\cite{Qiunan2018}, where $\bm{m}=m(0, 0, 1)$.
  
  We introduce the Kane-Mele type spin-orbit coupling in the Kagome layer\cite{Kane2005,Guo2009} as given as
  \begin{align}
H_{\mathrm{so}}=-\rm{i} \it{t}_{\rm{so}} \sum_{< ij >\sigma\sigma'} \nu_{ij} d^{\dagger}_{i\sigma} \sigma^z_{\sigma\sigma'}d _{j\sigma'}.
\end{align}
Here $t_{\rm{so}}$ is the hopping amplitude 
and the sign $\nu_{ij}=\pm1$ depends
on the
orientation of the two nearest neighbor bonds. When an
electron traverses in going from site $j$ to $i$,
$\nu_{ij}=\pm1$
, if the electron makes a left (right) turn to get to the second
bond.
The substantial strength of spin-orbit coupling originates from the presence of Sn2 atoms at the center of hexagons in the original lattice structures of Co$_3$Sn$_2$S$_2$. 
The $d$ electrons in the hexagons are susceptible to the strong potentials from Sn2's nucleuses.

\begin{figure}[t]
\includegraphics[width=0.9\hsize]{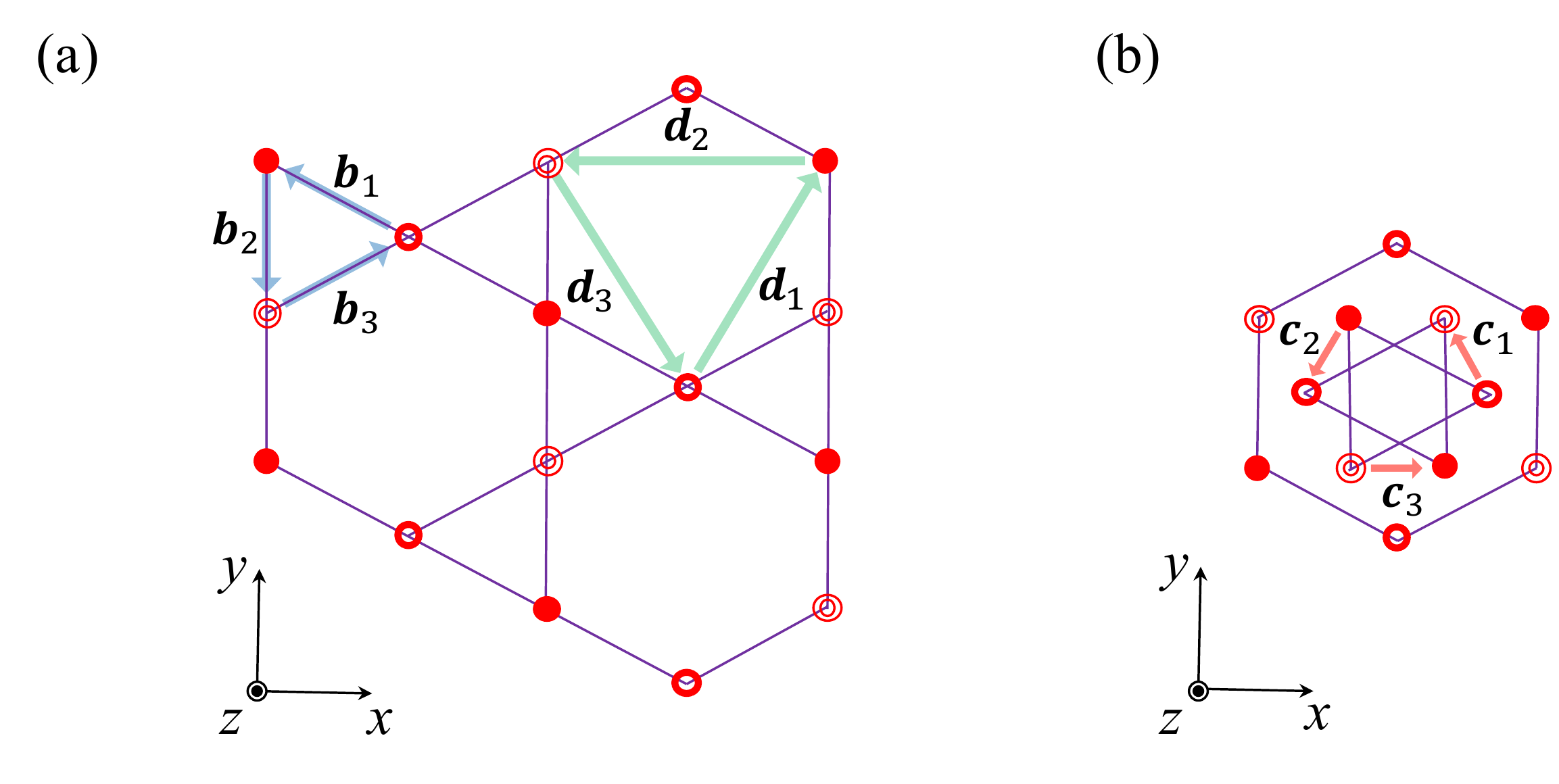}
\vspace{-3mm}
\caption{
(a)~Intralayer lattice vectors of a Kagome layer. 
$\bm{b}_1$, $\bm{b}_2$, $\bm{b}_3$ are the first-nearest-neighbor vectors.
$\bm{d}_1$, $\bm{d}_2$, $\bm{d}_3$ are the second-nearest-neighbor vectors.
(b)~Interlayer lattice vectors of Kagome layers. 
$\bm{c}_1$, $\bm{c}_2$, $\bm{c}_3$ are the first-nearest-neighbor vectors.
}
\label{fig:lattice}
\end{figure}

{\it Nodal rings and Weyl points}---
The Hamiltonian Eq.~(\ref{eq:hamiltonian}) is diagonalized by using the Fourier transformation $d_{i\sigma}=\frac{1}{\sqrt{N}} \sum_{\bm{k}} \mathrm{e}^{i\bm{k} \cdot \bm{R}_i}d_{\bm{k}\alpha\sigma}$ and $p_{i\sigma}=\frac{1}{\sqrt{N}} \sum_{\bm{k}} \mathrm{e}^{i\bm{k} \cdot \bm{R}_i}p_{\bm{k}\sigma}$.
Here $\bm{k}$ is the crystal momentum, and $\alpha=A,\ B, {\rm{or}}\ C$ is the sublattice index of the Kagome lattice.
For each $\bm{k}$ the Bloch wave function is an eight component eigenvector 
$\ket{u_{ n \bm{k} } }$
of the Bloch Hamiltonian matrix $ \mathcal{H}(\bm{k})$ which is given by
\begin{align}
H
&=\sum_{\bm{k},\sigma} C^{\dagger}_{\bm{k}\sigma} \mathcal{H}(\bm{k}) C_{\bm{k}\sigma},
\end{align}
where
$C_{\bm{k}\sigma}^{\dag}=(d^{\dag}_{\bm{k}A\sigma}, d^{\dag}_{\bm{k}B\sigma}, d^{\dag}_{\bm{k}C\sigma}, p^{\dag}_{\bm{k}\sigma})$.
$\mathcal{H}(\bm{k})$ consists of the following terms:
$\mathcal{H}(\bm{k})=\mathcal{H}_1+\mathcal{H}_2+\mathcal{H}_{\rm{dd}}+\mathcal{H}_{\rm{dp}}+\mathcal{H}_{\rm{p}}+\mathcal{H}_{\rm{exc}}+\mathcal{H}_{\rm{so}}$.
Here, each term is given as below.

\begin{align}
\mathcal{H}_1=-2t_1
\begin{pmatrix}
    0&      \cos(k^b_1)  \sigma_0&       \cos(k^b_3)\sigma_0&   0  \\
  \cos(k^b_1)  \sigma_0&         0&       \cos(k^b_2)\sigma_0&0  \\
   \cos(k^b_3) \sigma_0&        \cos(k^b_2)\sigma_0&          0&  0\\
      0 & 0& 0 & 0\nonumber
\end{pmatrix},
\end{align}

 \begin{align}
\mathcal{H}_2= -2t_2
\begin{pmatrix}
    0&      \cos(k^d_1) \sigma_0&       \cos(k^d_3) \sigma_0&   0  \\
  \cos(k^d_1)  \sigma_0&         0&       \cos(k^d_2) \sigma_0&  0\\
   \cos(k^d_3)  \sigma_0&        \cos(k^d_2)\sigma_0&          0&  0\\
   0 & 0& 0 & 0\nonumber
\end{pmatrix},
\end{align}
 
 \begin{align}
\mathcal{H}_{\rm{z}}=-2t_{\rm{z}}
\begin{pmatrix}
    0&      \cos(k^c_1)  \sigma_0&       \cos(k^c_3)\sigma_0&    0 \\
  \cos(k^c_1) \sigma_0&         0&       \cos(k^c_2)\sigma_0&   0  \\
   \cos(k^c_3) \sigma_0&        \cos(k^c_2) \sigma_0&          0& 0 \\
   0 & 0 & 0 &0 \nonumber
\end{pmatrix},
\end{align}

\begin{align}
\mathcal{H}_{\rm{dp}}=2\rm{i}\it{t}_{\rm{dp}}
\begin{pmatrix}
 0 & 0 & 0 & -\sin(k^a_1) \sigma_0\\
 0 & 0 & 0 & -\sin(k^a_2) \sigma_0\\
 0 & 0 & 0 & -\sin(k^a_3) \sigma_0\\
            \sin(k^a_1) \sigma_0 \!\!
           &\sin(k^a_2) \sigma_0 \!\!
           &\sin(k^a_3) \sigma_0 \!\!
           & 0\nonumber
\end{pmatrix},
\end{align}

\begin{align}
\mathcal{H}_{\rm{p}}=\epsilon_{\rm{p}}
\begin{pmatrix}
0&      0&     0 &     0        \\
  0&    0&    0 &        0     \\
  0&      0&     0&         0  \\
  0 &0&0&\sigma_0\nonumber
\end{pmatrix},
\end{align}

\begin{align}
\mathcal{H}_{\rm{exc}}=-J|\bm{m}|
\begin{pmatrix}
\sigma_z&      0&     0 &     0        \\
  0&    \sigma_z&    0 &        0     \\
  0&      0&     \sigma_z&         0  \\
  0 &0&0&0\nonumber
\end{pmatrix},
\end{align}

\begin{align}
\mathcal{H}_{\rm{so}}=-2\rm{i}\it{t}_{\rm{so}}
\begin{pmatrix}
    0&      -\cos(k^d_1) \sigma_z&       \cos(k^d_3)  \sigma_z& 0    \\
  \cos(k^d_1) \sigma_z&         0&       -\cos(k^d_2)\sigma_z& 0 \\
   -\cos(k^d_3)\sigma_z&        \cos(k^d_2)\sigma_z&          0& 0 \\
   0&0&0&0
\end{pmatrix}.
\end{align}
Here,~
$k^b_i=\bm{k} \cdot \bm{b}_i$,~
$k^c_i=\bm{k} \cdot \bm{c}_i$,~
$k^d_i=\bm{k} \cdot \bm{d}_i$,~
and $k^a_i=\bm{k} \cdot \bm{a}_i/2$,
~$i=1,2,3$.
These lattice vectors are shown in Fig.~\ref{fig:electrons}(b), Fig.~\ref{fig:lattice}(a) and Fig.~\ref{fig:lattice}(b).\par
\begin{figure}[t]
\includegraphics[width=85mm]{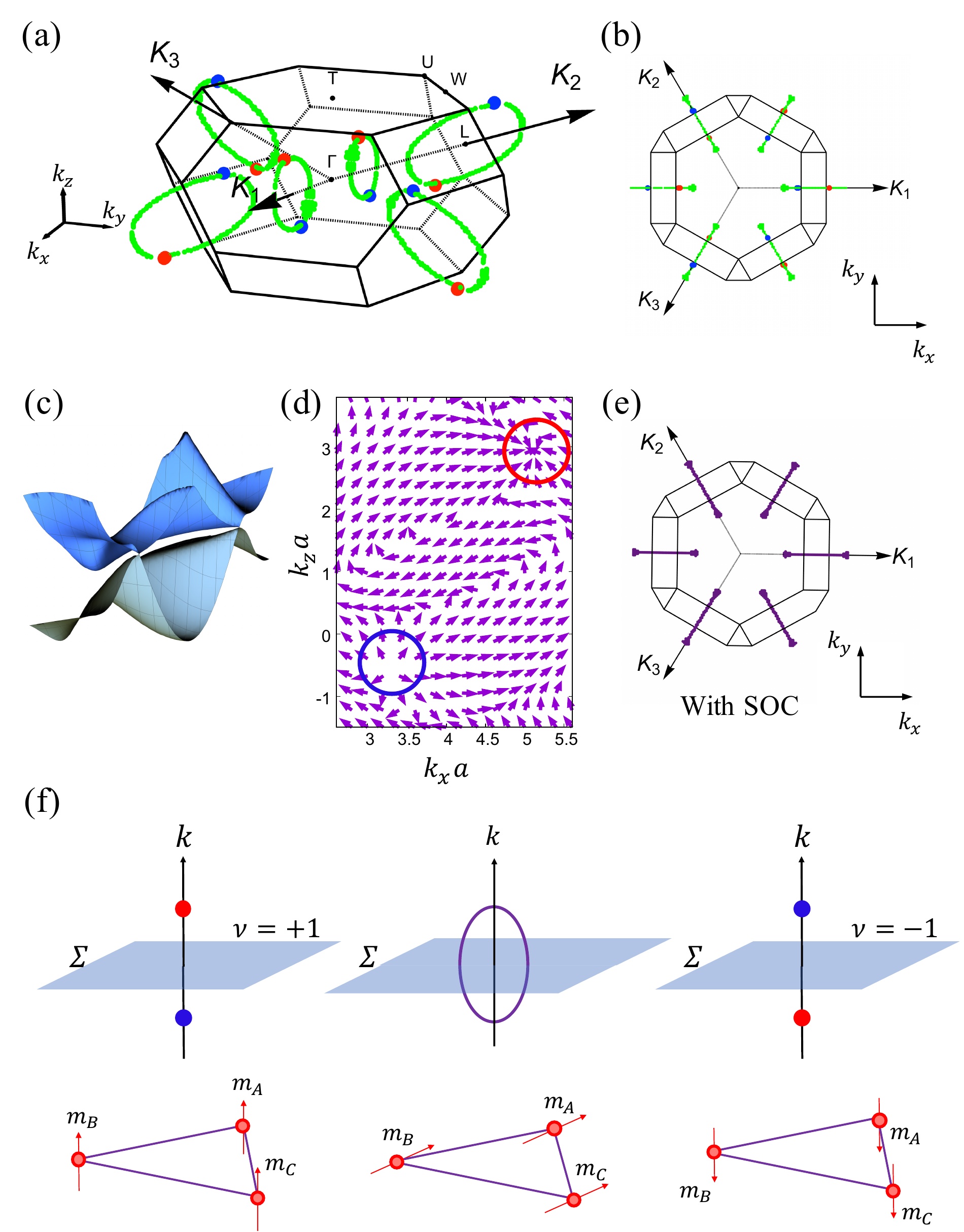}
\caption{
(a)(b)~Band touching points between $n=3$ and $n=4$ band without SOC~(green lines) and with SOC~(red and blue points). 
(c)~Energy spectrum in $k_y=0$ plane. Gapless linear dispersions appear near the Fermi level. 
(d)~Berry curvature distribution in the $k_y=0$ plane. The sink and source of the Berry curvature correspond to the Weyl points.
(e)~Nodal rings with spin-orbit coupling when the magnetization is perpendicular to the $z$-axis.
(f)~The relation between the Weyl points configuration and Chern number defined in the plane $\it{\Sigma}$. Chern number changes when the magnetization flips.
}
\label{fig:nodes}
\end{figure}

By solving the eigenvalue equation $\mathcal{H}(\bm{k})\ket{u_{ n \bm{k} } }=E_{n \bm{k}}\ket{u_{ n \bm{k} } }$, we obtain eigenstates $\ket{u_{ n \bm{k} } }$ and eigenvalues $E_{n \bm{k}}$, where $n$ (from 1 to 8) being the band index labeled from the bottom. 
\begin{figure}[t]
%\raggedleft
\includegraphics[width=1.0\hsize]{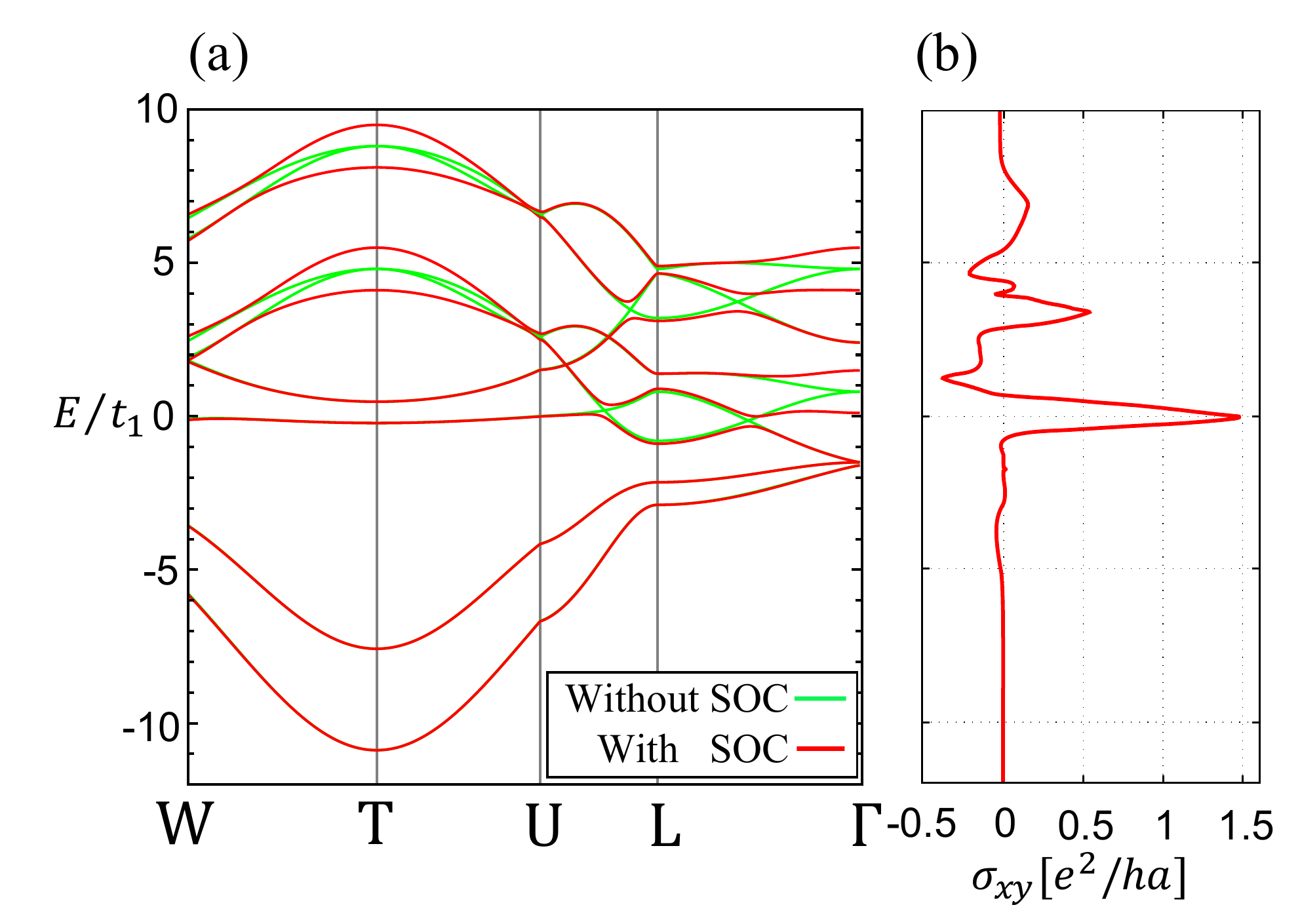}
\vspace{-7mm}
\caption{
(a)~Band structure on high symmetry lines. 
There is a band inversion between the $n=3$ band and $n=4$ band in the absence of spin-orbit coupling~(green lines). 
Spin-orbit coupling opens energy gaps on high symmetry lines~(red lines)
(b)~Energy dependence of the anomalous Hall conductivity. 
A peak structure near the energy of the Weyl points is observed.\label{fig:1}}
\label{fig:spa}
\end{figure}
The energy eigenvalues of the system calculated without and with spin-orbit coupling along high-symmetry lines are shown in Fig.~\ref{fig:spa}(a).
 Here we set $t_1$ as a unit of energy, $t_{2}=0.6t_{1}$, $t_{\rm{dp}}=1.0t_{1}$, $t_{\rm{z}}=-1.0t_{1}$,  $\epsilon_{\rm{p}}=-3.5t_{1}$, $J=2.0t_{1}$.
 We focus on $n=3$ band and $n=4$ band crossing the Fermi level.
When spin-orbit coupling is absent, nodal rings between the $n=3$ band and $n=4$ band appear around the L point. The positions of the nodal rings in momentum space are shown as green lines in Figs.~\ref{fig:nodes}(a) and \ref{fig:nodes}(b).
The above hopping parameters were chosen so that the configurations of the nodal rings are similar to those obtained by first principle calculations\cite{Qiunan2018}.

The nodal rings appearing in the absence of spin-orbit coupling are gapped out in the presence of spin-orbit coupling except two points on each ring. 
The energy spectrum shown in Fig.~\ref{fig:nodes}(c) is linear around the band touching points, which is consistent with the result of first principle calculations\cite{Enke2018}.
To characterize these nodal points we calculate the Berry curvature \cite{Xiao2010} $\bm{b}_{n \bm{k}}=\bm{\nabla} \times  \bm{a}_{n \bm{k}}$ of the $n=3$ band. 
Here $\bm{a}_{n \bm{k}}=-\rm{i}  \it{\braket{ u_{ n \bm{k} }|\bm{\nabla}_{\bm{k}}|u_{ n \bm{k} }}}$ is the Berry connection \cite{Xiao2010}. 
Figure \ref{fig:nodes}(d) shows the Berry curvature distribution in the  $k_y=0$ plane. 
There are sources and sinks of the Berry curvature corresponding to the Weyl points with positive chirality and negative chirality, respectively.

Next, we examine the intrinsic anomalous Hall conductivity by using the Kubo formula\cite{Nagaosa2010},
\begin{align}
\sigma_{xy}= \frac{e^2}{h} \sum_n \int_{\mathrm{BZ}} \frac{d^3\bm{k}}{(2 \pi)^2} b^z_{n\bm{k}}f(E_{n{\bm{k}}}-\mu).
\label{eq:kubo}
\end{align}
Here, $n$ is the occupied band index, $b^z_{n\bm{k}}$ is the $z$ component of the Berry curvature,  $f$ is the Fermi-Dirac distribution function and $\mu$ is the Fermi level.
Figure~\ref{fig:spa}(b) shows the Fermi level dependence of the anomalous Hall conductivity. 
Near the energy of the Weyl points,  the anomalous Hall conductivity has a large  peak. 
The value near the Fermi energy $\sigma_{xy} \approx 1059 \Omega^{-1} \rm{cm^{-1}}$ is very close to the result of first principle calculation and experiment\cite{Enke2018,Qi2018}.
In an ideal Weyl semimetal where Fermi surfaces reside only at the Weyl points,
we can compute the anomalous Hall conductivity as the summation of the distance of the Weyl points separated by  $\Delta K_{z}^{(\gamma)}$ as following \cite{Wan2011,Burkov2011,Armitage2018},
\begin{align}
\sigma_{xy}^{\rm{Weyl}}= \frac{e^2}{2 \pi h}  \sum_{\gamma} \Delta K_{z}^{(\gamma)}. 
\label{eq:distance}
\end{align}
Here, $\gamma$ indicates the pair of the Weyl points. 
The value of the anomalous Hall conductivity computed by Eq.~(\ref{eq:distance}) is $ \sigma_{xy} \approx 0.86 [e^2/ha]$.
This value is in reasonable agreement with that at the energy of the Weyl points computed by Eq.~(\ref{eq:kubo}) with our model, $\sigma_{xy} \approx 0.61 [e^2/ha]$.  

In the above calculation, we showed that the Weyl points appear in the presence of  spin-orbit coupling when the magnetization is parallel to the $z$-axis.
Here, we study the energy spectrum when the magnetization is perpendicular to the $z$-axis.
In this situation, nodal rings appear even in the presence of spin-orbit coupling.
In-plane magnetizations are set as $\bm{m}_A=\bm{m}_B=\bm{m}_C=m\left( 1,  0  ,0    \right)$.
~Figure~\ref{fig:nodes}(e) shows nodal rings between the $n=3$ band and $n=4$ band which correspond to nodal rings in Fig.~\ref{fig:nodes}(b).  
This appearance of nodal rings with spin-orbit coupling can be understood in terms of the Chern number.
We consider a certain plane in momentum space as shown in Fig.~\ref{fig:nodes}(f).
For simplicity, we consider a system with two Weyl points\cite{Armitage2018}.
The Hamiltonian in this plane, say $\it{\Sigma}$, can be regarded as that of a two-dimensional quantum anomalous Hall state\cite{Armitage2018}.
The Chern number  of $n\rm{th}$ band, $\nu_n(\it{\Sigma})=\int_{\Sigma} b_{n}^z($$\bm{k}$$) d$$\bm{k}$, can be defined in this plane $\it{\Sigma}$, if the plane does not contain any Weyl points\cite{Armitage2018}.
The sign of the Chern number changes, when it is finite, with the flip of the magnetization.
The change of the Chern number occurs only when the band gap closes.
Therefore, when the magnetization flips from one direction to the opposite direction, nodal rings need to appear as shown in Fig.~\ref{fig:nodes}(f).

%\section{Weyl points}

\begin{figure}[t] 
\includegraphics[width=1\hsize]{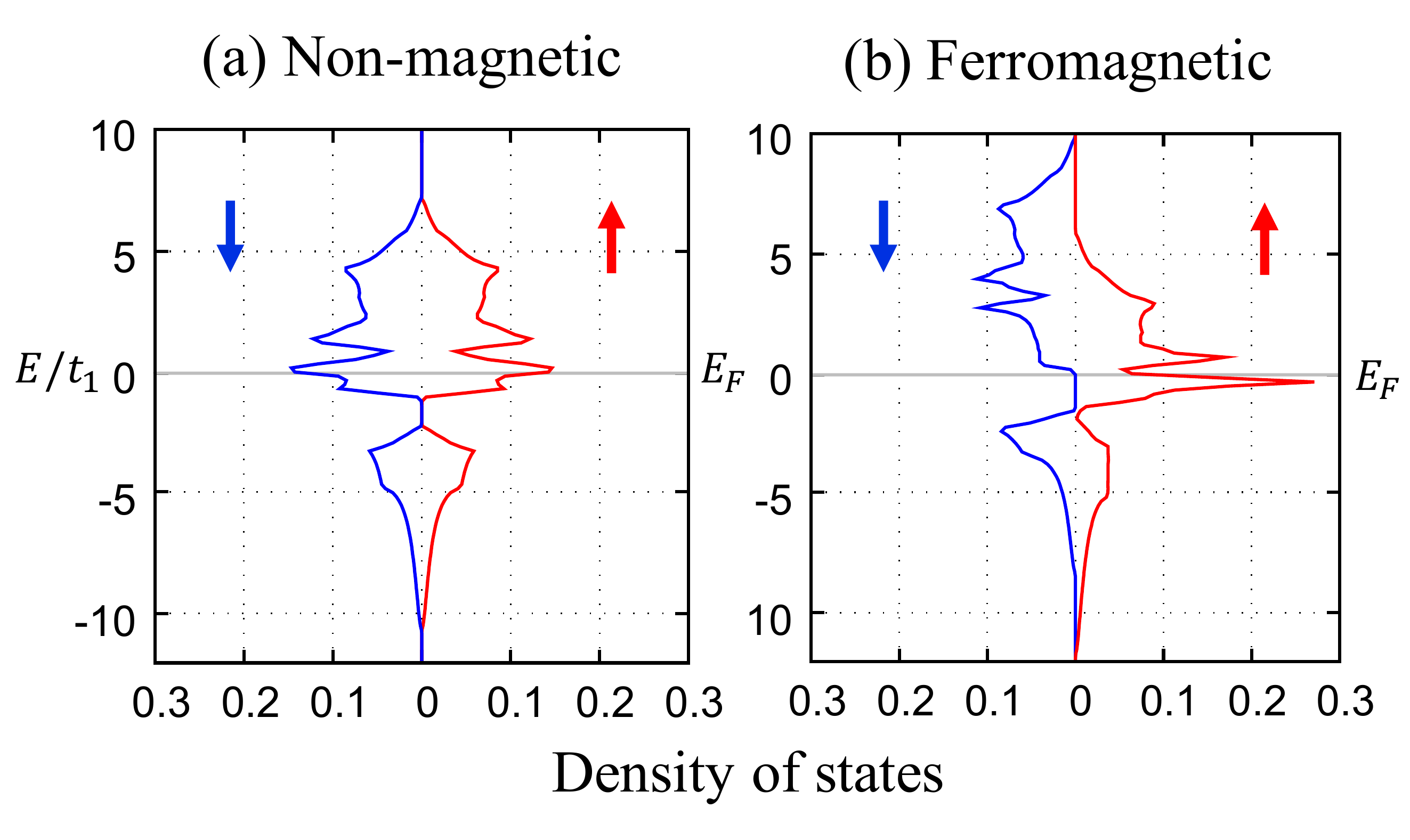}
\vspace{-7mm}
\caption{The density of states of (a)~non-magnetic state $J|\bm{m}|=0$. A peak near the Fermi level stabilizes  ferromagnetic order and of (b)~ferromagnetic state $J|\bm{m}|\not = 0$. A minimum of the density of states near the Fermi level is shown.}
\label{fig:dos}
\end{figure}

{\it Magnetic order}---
Next, we discuss the origin of magnetism in shandite materials based on the Stoner theory\cite{Jurgen2017}.
Although most shandite materials are non-magnetic, $\rm{{Co}_3{Sn}_2{S}_2}$ shows ferromagnetic order\cite{Weihrich2004,Umetani2008}.
Additionally, this ferromagnetism is suppressed by substituting, for example, Co for Fe or Ni, Sn for In\cite{Weihrich2006,Kassem2015,Kassem2016jc,Kassem2016jp,Kubodera2006,Sakai2015}.
According to the Stoner theory, ferromagnetism of itinerant electrons is characterized by the following criterion\cite{Jurgen2017},
\begin{align}
D(E_F)U>1.
\label{eq:Stoner}
\end{align}
Here, $D(E_F)$ is the density of states in non-magnetic state at the Fermi level and 
$U$ is the onsite Coulomb interaction.
In order to examine this criterion, we calculate the density of states in non-magnetic state as shown in Fig.~\ref{fig:dos}(a).
The Fermi level can be calculated by the 3/8 filling condition.
The density of states has a peak structure near the Fermi level.
This peak is significant to satisfy the Stoner criterion Eq.~(\ref{eq:Stoner}).
When the number of electrons changes due to the chemical substituent, the Fermi level shifts from the peak and the density of states decreases, suppressing ferromagnetic order. 
Contrary, in ferromagnetic state, the density of states has a minimum near the Fermi level as shown in Fig.~\ref{fig:dos}(b).

%\section{Magnetic anisotropy}

\begin{figure}[t]
\includegraphics[width=1.0\hsize]{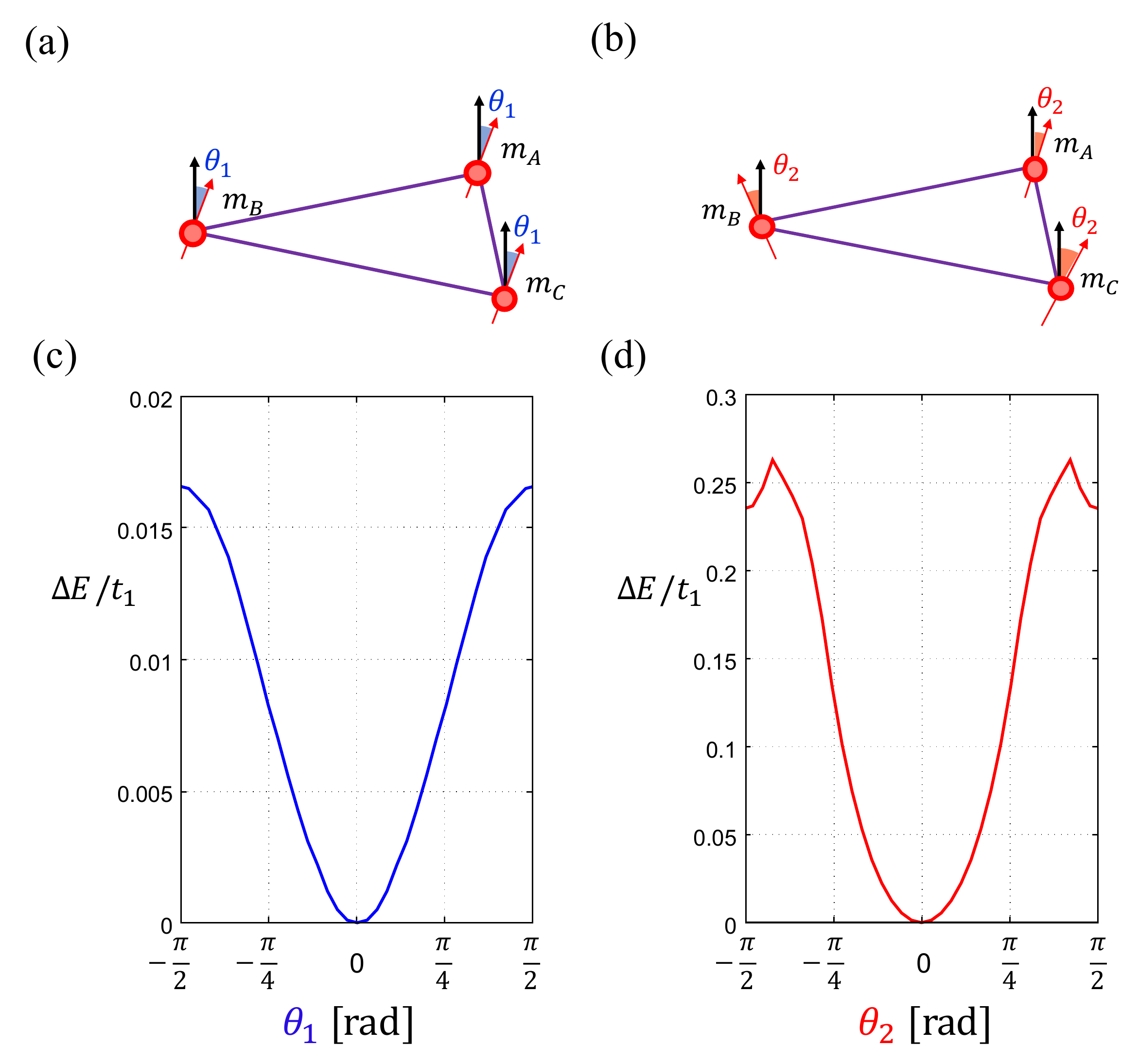}
\vspace{-7mm}
\caption{
(a)~Uniformly tilted magnetizations.
(b)~Umbrella structure of magnetizations.
(c)~(d)~Angle dependence of the total energy. Magnetizations pointing along the $z$-axis, $\theta_1=\theta_2=0$, are energetically favored.}	
	\label{fig:aniso}
\end{figure}

{\it Magnetic anisotropy}---
We examine the easy-axis anisotropy of our model.
We calculate the magnetization-angle dependence of the total energy with two-types of tilted configurations.
In the first case, magnetizations on each sublattice are given as $\bm{m}_A\!=\!\bm{m}_B\!=\!\bm{m}_C\!=\!\left( \sin\theta_1,  0  ,\cos \theta_1    \right)$ as shown in Fig.~\ref{fig:aniso}(a).
In the second case, magnetizations on each sublattice are given as 
$\bm{m}_A\!=\!m \left( \sin\theta_2,  0  ,\cos \theta_2    \right)$, 
$\bm{m}_B\!=\!m(  -\frac{1}{2}\sin\theta_2, \frac{\sqrt{3}}{2}\sin\theta_2 , \cos \theta_2   )$, 
$\bm{m}_C\!=\!m(  -\frac{1}{2}\sin\theta_2, -\frac{\sqrt{3}}{2}\sin\theta_2 , \cos \theta_2   )$ 
as shown in Fig.~\ref{fig:aniso}(b).
Here, $\theta_1$ and $\theta_2$ are tilting angles.
The first tilting case Fig.~\ref{fig:aniso}(a) corresponds to the situation when the applied magnetic filed points in the direction perpendicular to the $z$-axis.
The second tilting case Fig.~\ref{fig:aniso}(b) corresponds to  umbrella structure of the magnetization suggested by experiment\cite{Kassem2017}.
We compute the total energy of electrons, $E=\frac{1}{N}\sum_{n,\bm{k}} E_{n\bm{k}} f(E_{n\bm{k}}-\mu)$ as a function of $\theta_1$ and $\theta_2$ with two cases of magnetizations Figs.~\ref{fig:aniso}(a) and \ref{fig:aniso}(b).
Here $N$ is the number of the unit cells and $\mu$ is the Fermi level calculated under the 3/8 filling condition at each $\theta_1$, $\theta_2$.
Figures~\ref{fig:aniso}(c) and \ref{fig:aniso}(d) show the energy shifts from the total energy with out-of-plane magnetization as functions of $\theta_1$ and $\theta_2$, respectively.
In both case, the total energy has a minimum at $\theta_1=\theta_2=0$.
This behavior shows the easy-axis ferromagnetic anisotropy, which is consistent with experiment\cite{Lin2012,Schnelle2013}.\par

%\section{conclusion}

{\it Conclusion}---
In this work, we constructed an effective tight binding model for the ferromagnetic Co-shandite $\rm{{Co}_3{Sn}_2{S}_2}$. 
The configurations of nodal rings and the Weyl points are similar to those obtained by first principle calculations\cite{Enke2018,Qiunan2018}. 
When the magnetization is perpendicular to the $z$-axis,  nodal rings appear even with spin-orbit coupling.
We showed that this model describes the itinerant magnetism and easy-axis anisotropy of $\rm{{Co}_3{Sn}_2{S}_2}$, which are consistent with experiment\cite{Weihrich2004,Umetani2008,Weihrich2006,Kassem2015,Kassem2016jc,Kassem2016jp,Kubodera2006,Sakai2015,Lin2012,Schnelle2013}.

{\it Acknowledgement}---
I would like to thank
Y.~Araki,
J.~Checkelsky,
K.~Kobayashi,
T.~Koretsune,
D.~Kurebayashi,
Y.~Motome,
Y.~Nakamura,
M.-T.~Suzuki,
A.~Tsukazaki,
A.~Yamakage, and
Y.~Yanagi for helpful discussions.
This work was supported by 
JSPS KAKENHI Grants No. JP15H05854 and No. JP17K05485,
JST CREST Grant No. JPMJCR18T2, 
and GP-Spin at Tohoku University.


\begin{thebibliography}{99}

\bibitem{Wan2011}
X. Wan, A. M. Turner, A. Vishwanath, and S. Y. Savrasov, Phys. Rev. B  {\bf83}, 205101 (2011).

\bibitem{Burkov2011}
A. A. Burkov and L. Balents, Phys. Rev. Lett. {\bf107}, 127205 (2011).

\bibitem{Armitage2018}
N. P. Armitage, E. J. Mele, and A. Vishwanath, Rev. Mod. Phys. {\bf90}, 015001 (2018).

\bibitem{Xiao2010}
D. Xiao, M.-C. Chang, and Q. Niu, Rev. Mod. Phys. {\bf 82}, 1959 (2010).

\bibitem{Nagaosa2010}
N. Nagaosa, J. Sinova, S. Onoda, A. MacDonald, and N. Ong, Rev. Mod. Phys. {\bf 82}, 1539 (2010).

\bibitem{Nomura2015}
K. Nomura and D. Kurebayashi, Phys. Rev. Lett. {\bf 115}, 127201 (2015).
%\label{Nomura2015Ref}

\bibitem{Kurebayashi2016}
D. Kurebayashi and K. Nomura, Phys. Rev. Applied  {\bf6}, 044013 (2016).

\bibitem{Araki2016}
Y. Araki,  A. Yoshida, and K. Nomura, Phys. Rev. B {\bf 94}, 115312 (2016).

\bibitem{Kurebayashi2019}
D. Kurebayashi and K. Nomura, Scientific Reports {\bf8}, 5365 (2019).

\bibitem{Chen2014}
H. Chen, Q. Niu, and A. H. MacDonald, Phys. Rev. Lett. {\bf 112}, 017205 (2014).

\bibitem{Kubler2014}
J. K\"ubler and C. Felser, Europhys, Lett. {\bf 108}, 67001 (2014).

\bibitem{Nakatsuji2015}
S. Nakatsuji, N. Kiyohara, and T. Higo, Nature {\bf 527}, 212 (2015).

\bibitem{Yang2017}
H. Yang, Y. Sun, Y. Zhang, W.-J. Shi, S. S. P. Parkin, and B. Yan, New J. Phys. {\bf 19} 015008 (2017).

\bibitem{Suzuki2017}
M.-T. Suzuki, T. Koretsune, M. Ochi, and R. Arita, Phys. Rev. B {\bf 95}, 094406 (2017).


\bibitem{Ito2017}
N. Ito and K. Nomura, J. Phys. Soc. Jpn. {\bf 86}, 6, 063703 (2017).

\bibitem{Jianpeng2017}
J. Liu and L. Balents, Phys. Rev. Lett. {\bf119}, 087202 (2017).

\bibitem{Kuroda2017}
K. Kuroda, T. Tomita, M.-T. Suzuki, C. Bareille, A. A. Nugroho, P. Goswami, M. Ochi, M. Ikhlas, M. Nakayama, S. Akebi, et al,~Nature Materials, {\bf16}, 11, 1090 (2017).

\bibitem{Zhijun2016}
Z. Wang, M. G. Vergniory, S. Kushwaha, M. Hirschberger, E. V. Chulkov, A. Ernst, N. P. Ong, R. J. Cava, and B. A. Bernevig, Phys. Rev. lett. {\bf117}, 23, 236401 (2016).


\bibitem{Guoqing2016}
G. Chang, S.-Y. Xu, H. Zheng, B. Singh, C.-H. Hsu, G. Bian, N. Alidoust, I. Belopolski, D. S. Sanchez, S. Zhang, et al, Scientific Reports {\bf6}, 38839 (2016).

\bibitem{Enke2018}
E. Liu, Y. Sun, N. Kumar, L. Muechler, A. Sun, L. Jiao, S.-Y. Yang, D. Liu, A. Liang, Q. Xu, et al, Nat. Phys. {\bf14}, 1125-1131 (2018).

\bibitem{Qiunan2018}
Q.~Xu, E.~Liu, W.~Shi, L.~Muechler, J.~Gayles, C.~Felser, and Y.~Sun, Phys. Rev. B {\bf97}, 235416 (2018).


\bibitem{Qi2018}
Q. Wang, Y. Xu, R. Lou, Z. Liu, M. Li, Y. Huang, D. Shen, H. Weng, S. Wang, and H. Lei, Nat. Comm. {\bf9}, 1, 3681 (2018).

\bibitem{Kubodera2006}
T. Kubodera, H. Okabe, Y. Kamihara, and M. Matoba, Phys. B Condens. Matter {\bf378-380}, 1142 (2006).

\bibitem{Sakai2015}
Y. Sakai, R. Tanakadate, M. Matoba, I. Yamada, and N. Nishiyama, J. Phys. Soc. Jpn. {\bf84}, 44705 (2015).

\bibitem{Paz2009}
P. Vaqueiro and G. G. Sobany, Solid State Sci. {\bf11}, 513-518 (2009).

\bibitem{Jurgen2017}
Ju\"rgen Ku\"bler. {\it Theory of itinerant electron magnetism}, (Oxford University Press, Oxford) (2000).

\bibitem{Kane2005}
C. L. Kane and E. J. Mele, Phys. Rev. Lett. {\bf95}, 226801, (2005).

\bibitem{Guo2009}
H.-M. Guo and M. Franz, Phys. Rev. B {\bf80}, 113102, (2009).


\bibitem{Weihrich2004}
R. Weihrich, I. Anusca, and M. Zabel, Z. Anorg. Allg. Chem. {\bf630}, 1767 (2004).

\bibitem{Weihrich2006}
R. Weihrich and I. Anusca, Z. Anorg. Allg. Chem. {\bf632}, 1531 (2006).

\bibitem{Umetani2008}
A. Umetani, E. Nagoshi, T. Kubodera, and M. Matoba, Phys. B Condens. Matter {\bf403}, 1356 (2008).


\bibitem{Lin2012}
X. Lin, S. L. Bud’ko, and P. C. Canfield, Philos, Mag. {\bf92}, 2436 (2012).

\bibitem{Schnelle2013}
W. Schnelle, A. Leithe-Jasper, H. Rosner, F. M. Schappacher, R. Po\"ttgen, F. Pielnhofer, and R.
Weihrich, Phys. Rev. B {\bf88}, 144404 (2013).


\bibitem{Kassem2015}
M. A. Kassem, Y. Tabata, T. Waki, and H. Nakamura, J. Cryst. Growth {\bf426}, 208 (2015).


\bibitem{Kassem2016jc}
M. A. Kassem, Y. Tabata, T. Waki, and H. Nakamura, J. Solid State Chem. {\bf233}, 8 (2016).

\bibitem{Kassem2016jp}
M. A. Kassem, Y. Tabata, T. Waki, and H. Nakamura, J. Phys. Soc. Jpn. {\bf85}, 064706 (2016).

\bibitem{Kassem2017}
M. A. Kassem, Y. Tabata, T. Waki, and H. Nakamura,
Phys. Rev. B {\bf96}, 014429 (2017).


\end{thebibliography}
\end{document}